\documentclass[twocolumn,aip,reprint,superscriptaddress]{revtex4-2} 

\usepackage[T1]{fontenc}
\usepackage{amsmath,amsfonts,bm}

\usepackage{bbm}

\usepackage{amssymb}
\usepackage{graphicx}  
\usepackage{color}
\usepackage{extarrows}
\usepackage{enumitem}
\usepackage{empheq}
\usepackage{tensor}
\usepackage{orcidlink}
\usepackage[ruled, vlined]{algorithm2e}

 \usepackage[scr=boondoxupr]{mathalfa} 

\newcommand{\ud}{{\mathrm d}}

\newcommand{\B}{\mbox{\tiny B}}

\newcommand{\tS}{\mbox{\tiny S}}
\newcommand{\T}{\mbox{\tiny T}}

\newcommand{\SB}{\mbox{\tiny SB}}
\newcommand{\dg}{\dagger}
\newcommand{\la}{\langle}
\newcommand{\ra}{\rangle}

\newcommand{\Sec}[1]{Sec.\,\ref{#1}}
\newcommand{\App}[1]{Appendix\,\ref{#1}}
\renewcommand{\nl}{\nonumber \\}
\newcommand{\be}{\begin{equation}}
\newcommand{\ee}{\end{equation}}
\newcommand{\bsube}{\begin{subequations}}
\newcommand{\esube}{\end{subequations}}
\newcommand{\Eq}[1]{Eq.\,(\ref{#1})}
\newcommand{\Eqs}[1]{Eqs.\,(\ref{#1})}
\newcommand{\Fig}[1]{Fig.\,\ref{#1}}

\newcommand{\RN}[1]{%
  \textup{\uppercase\expandafter{\romannumeral#1}}%
}

\usepackage{siunitx}

\allowdisplaybreaks[1]

\definecolor{darkblue}{RGB}{0, 56, 102}

\usepackage{hyperref}  
\hypersetup{hidelinks,
	colorlinks=true,
	allcolors=black,
	pdfstartview=Fit,
	breaklinks=true
}

\begin{document}

\title{Fermionic influence superoperator for transport through Majorana zero modes
}

\author{Jia-Lin Pan}
\thanks{These authors contributed equally to this work.}
\affiliation{
  Department of Chemical Physics, University of Science and Technology of China, Hefei, Anhui 230026, China
}
\affiliation{
  Hefei National Research Center for Physical Sciences at the Microscale, University of Science and Technology of China, Hefei, Anhui 230026, China
}

\author{Zi-Fan Zhu}
\thanks{These authors contributed equally to this work.}
\affiliation{
  Hefei National Research Center for Physical Sciences at the Microscale, University of Science and Technology of China, Hefei, Anhui 230026, China
}

\author{Shixuan Chen}
\affiliation{
  Department of Chemical Physics, University of Science and Technology of China, Hefei, Anhui 230026, China
}
\affiliation{
  Hefei National Research Center for Physical Sciences at the Microscale, University of Science and Technology of China, Hefei, Anhui 230026, China
}

\author{Yu Su}
\email{suyupilemao@mail.ustc.edu.cn}
\affiliation{
  Hefei National Research Center for Physical Sciences at the Microscale, University of Science and Technology of China, Hefei, Anhui 230026, China
}

\author{Yao Wang}
\email{wy2010@ustc.edu.cn}
\affiliation{
  Hefei National Research Center for Physical Sciences at the Microscale, University of Science and Technology of China, Hefei, Anhui 230026, China
}


\date{\today}

\begin{abstract}
  In recent years, the study of Majorana signatures in quantum transport has become a central focus in condensed matter physics. Here, we present a rigorous and systematic derivation of the fermionic superoperator describing the open quantum dynamics of electron transport through Majorana zero modes, building on the techniques introduced in Phys.\,Rev.\,B {\bf 105}, 035121 (2022). The numerical implementation of this superoperator is to construct its differential equivalence, the hierarchical equations of motion (HEOM). The HEOM approach describes the system-bath correlated dynamics.
  Furthermore, we also develop a functional derivative scheme that provides exact expressions for the transport observables in terms of the auxiliary density operators introduced in the HEOM formulation. The superoperator formalism establishes a solid theoretical foundation for analyzing key transport signatures that may uncover the unique characteristics of Majorana physics in mesoscopic systems.
\end{abstract}

\maketitle

\section{Introduction}
The search for Majorana zero modes (MZMs)---quasiparticles with non-Abelian statistics---has become a major frontier in condensed matter physics, driven by their potential applications in topological quantum computation and fault-tolerant quantum information processing.\cite{Mou121003,Nad14602,Kit01131,Iva01268} These exotic states, localized at the ends of one-dimensional topological superconductors, exhibit robustness against local perturbations and are expected to manifest unique signatures in quantum transport experiments, such as zero-bias conductance peaks and fractional Josephson effects.\cite{Fu08096407,Kwo04349}

Recent advances in hybrid nanostructures, including semiconductor nanowires with strong spin-orbit coupling and proximity-induced superconductivity, have enabled experimental observations consistent with MZMs.\cite{Mou121003,Nad14602} However, distinguishing genuine Majorana signatures from trivial Andreev bound states remains a critical challenge.\cite{Liu11201308,Pra20575,Jin22093009,Cao12115311,Zaz11165440,Fle10180516} Theoretical frameworks that accurately describe open quantum dynamics in these systems are thus essential for interpreting transport data and guiding future experiments.\cite{Che19057403}

In this work, we develop a canonical fermionic superoperator formalism to model the nonequilibrium quantum transport through MZMs, building on the operator-space techniques established in Ref.\,[\onlinecite{Cir22035121}]. We express the environmental influence on the system in terms of a fermionic influence superoperator, where the bath properties are fully characterized by their two-time correlation functions. For numerical solving the influce superoperator, we construct its differential equivalence, the hierarchical equations of motion (HEOM) formalism. The HEOM approach describes the dynamics of the reduced density operator and a set of auxiliary density operators (ADOs), which encode the system-bath correlations.\cite{Tan89101,Tan906676,Yan04216,Xu05041103,Yan14054105,Yan16110306,Su24084104} Beyond Ref.\,[\onlinecite{Cir22035121}], we propose the functional derivative scheme on the influence superoperator, which allows us to derive exact expressions for transport observables in terms of the ADOs. Further generalizations of our scheme lead to the inner relations among the ADOs. This lays the foundation for the concept of the statistical quasi-particle, dissipaton, proposed in Ref.\,[\onlinecite{Yan14054105}]. As a result, our approach provides a rigorous foundation for analyzing characteristics in Majorana-based devices, offering insights beyond Markovian treatments. 

The paper is organized as follows. In \Sec{sec2}, we introduce the theoretical model. In \Sec{sec3}, we present a canonical derivation of the fermionic influence superoperator governing the transport through MZMs. Then we construct the differentiate equivalence of the influence functional, the HEOM in \Sec{sec4}. Furthermore, we derive the formula for evaluating the transport quantities via the HEOM. Numerical demonstrations are carried out in \Sec{sec5}. Finally, in \Sec{sec6}, we summarize our results and outline potential future work.
Throughout this paper, we set $\hbar=1$ and $\beta_\alpha=1/(k_BT_{\alpha})$, with $k_B$ being the Boltzmann constant and $T_{\alpha}$ being the temperature of $\alpha$-lead (with $\alpha = \rm L$ representing the left one and $\alpha = \rm R$ for the right one). 

\section{Theoretical models}\label{sec2}

\subsection{Model Hamiltonian}

The transport setup is the same as that in Ref.\,[\onlinecite{Jin22093009}]. The total system-bath Hamiltonian reads
\be \label{HT}
H_{\T}=H_{\tS}+H_{\SB}+\tilde h_{\B}.
\ee
Here, the central system consists of a pair of MZMs, 
\be \label{Hs}
H_{\tS}=\frac{i}{2}\varepsilon_{\rm M}\hat \gamma_{\rm L}\hat \gamma_{\rm R},
\ee
where $\hat \gamma_{\rm L/R}$ are two MZMs, satisying $\hat \gamma^{\dg}_{\rm L/R}=\hat \gamma_{\rm L/R}$, coupled to each other with  coupling constant $\varepsilon_{\rm M}$.
The bath Hamiltonian reads 
\be \label{hb}
h_{\B}= \sum_{\alpha} h_{\alpha} = \sum_{\alpha k}\varepsilon_{\alpha k}\hat c_{\alpha k}^{\dg}\hat c_{\alpha k},\ \ \ \ (\alpha={\rm L}\ \text{and}\ {\rm R}),
\ee
with $\hat c_{\alpha k}^{\dg}$ and $\hat c_{\alpha k}$ being the creation and annihilation operators of the $k$-state electron in the $\alpha$-lead. For imposing the current transport, we apply the electric potential $\varphi_{\alpha}$ on each lead, leading to 
\begin{align}
  \tilde h_{\B} = h_{\B} + \sum_{\alpha}\varphi_{\alpha}\hat N_{\alpha}
\end{align}
with $\hat N_{\alpha} \equiv \sum_{k}\hat c_{\alpha k}^\dagger\hat c_{\alpha k}$ being the bath particle number operator. The system-bath interaction reads
\begin{align}\label{Hsb}
H_{\SB}=&\sum_k \big[(t_{{\rm L}k} \hat c^{\dg}_{{\rm L}k} \hat \gamma_{\rm L}+t^{\ast}_{{\rm L}k} \hat \gamma_{\rm L}\hat c_{{\rm L}k})
\nl &
+i(t_{{\rm R}k} \hat c_{{\rm R}k}^{\dg} \hat \gamma_{\rm R}-t_{{\rm R}k}^{\ast}\hat \gamma_{\rm R}\hat c_{{\rm R}k})\big].
\end{align}

\subsection{Transformation to the regular-fermions representation}

The Majorana fermions can be expressed as
\be \label{regular} 
\hat \gamma_{{\rm L}}=\hat f+{\hat f}^{\dg}\ \ \text{and}\  \ \hat\gamma_{{\rm R}}=-i(\hat f-{\hat f}^{\dg}).
\ee
Here, $\hat f$ and $\hat f^{\dg}$ are regular fermions, satisfying $\{\hat f, \hat f^{\dg}\}=1$ and $\hat f^2=(\hat f^{\dg})^2=0$.
Equation (\ref{regular}) reproduces the properties of Majoranan fermions,
\be 
\{\hat\gamma_{\alpha},\hat\gamma_{\alpha'}\}=2\delta_{\alpha\alpha'},
\ee
and
\be 
\hat \gamma^{\dg}_{\alpha}=\hat \gamma_{\alpha}\ \ \text{and}\ \ \hat\gamma_{\alpha}^2=1.
\ee

The system Hamiltonian in \Eq{Hs} can then be recast as
\be \label{hs2}
H_{\tS}=\varepsilon_{\rm M}(\hat f^{\dg}\hat f-1/2).
\ee
And the system-bath interaction Hamiltonian in \Eq{Hsb} is reformulated as
\be \label{hsb2}
H_{\SB}=\sum_{\alpha} (\hat s_{\alpha}\hat F_{\alpha}^{\dg}+\hat F_{\alpha}\hat s_{\alpha}^{\dg})
\ee
with $\hat F_{\alpha}\equiv \sum_k t_{\alpha k}^{\ast}\hat c_{\alpha k}$, and
\begin{align}
\hat s_{\rm L}&\equiv -\gamma_{\rm L}=-(\hat f+\hat f^{\dg})=\hat  s_{\rm L}^{\dg},
\\
\hat  s_{\rm R}&\equiv -i\gamma_{\rm R}=-(\hat f-\hat f^{\dg})=-\hat  s_{\rm R}^{\dg}.
\end{align}
In the next section, we will give the canonical derivation of the influence superoperator based on \Eqs{hs2}, (\ref{hsb2}), and (\ref{hb}) based on the techniques established in Ref.\,[\onlinecite{Cir22035121}].


\section{Canonical derivation of the influence superoperator}\label{sec3}

\subsection{Graded tensor product of system and bath}

The quantum system-bath hybridization dynamics is described by quantum states living in the composite Hilbert space $\mathcal H_{\T} = \mathcal H_{\B} \times \mathcal H_{\tS}$. The total space operator generated by the system subspace operator $\hat A_{\tS}$ and bath one $\hat F_{\B}$ is defined via the tensor product $\hat F_{\B}\otimes\hat A_{\tS}$, satisfying the multiplication rule
\begin{align}
  (\hat F_{\B}\otimes\hat A_{\tS})(\hat G_{\B}\otimes\hat B_{\tS}) = \hat F_{\B}\hat G_{\B}\otimes\hat A_{\tS}\hat B_{\tS}.
\end{align}
However, for the fermionic operators, such a definition cannot lead to the anti-commutation relation between the system and bath operators, i.e., $\hat f\hat c_{\alpha k} = -\hat c_{\alpha k}\hat f$. Consequently, we have to define the graded tensor product $\otimes_g$, with \cite{Sch16155142,Sap12235432}
\begin{align}
  (\hat F_{\B}&\otimes_g\hat A_{\tS})(\hat G_{\B}\otimes_g\hat B_{\tS})\nl 
  &= (-)^{(\deg \hat A_{\tS})(\deg \hat G_{\B})}( \hat F_{\B}\hat G_{\B}\otimes_g\hat A_{\tS}\hat B_{\tS}).
\end{align}
Here, the degree of a system operator is defined as
\begin{align}
  \deg \hat A_{\tS} \equiv 
  \begin{cases}
    0, &\text{if $[\hat A_{\tS}, \hat P_{\tS}] = 0$}, \\
    1, &\text{if $\{\hat A_{\tS}, \hat P_{\tS}\} = 0$},
  \end{cases}
\end{align}
with $\hat P_{\tS} \equiv \exp(i\pi\hat f^\dagger\hat f)$ being the system parity operator. The definition for the bath operator is similar, also with the bath parity being 
\begin{align}
  \hat P_{\B} \equiv \prod_{\alpha k}\exp(i\pi\hat c_{\alpha k}^\dagger\hat c_{\alpha k}) = \exp\bigg(\sum_\alpha i\pi\hat N_{\alpha}\bigg).
\end{align}
Note that in defining the graded tensor product, we only consider those operators with definite parity. For an arbitrary operator, we can first decomposite it into the even (+1 parity) and odd ($-1$ parity) parts, and the proceed with the graded tensor product. See the detailed discussions on the parity in \App{AppA}.

However, for constructing the canonical representation of the influence functional, the graded tensor product presents additional algebraic complexities. Thus, we have to map the graded algebra into the normal one ($\otimes$). This is achieved by the extension
\begin{align}\label{ext}
  \mathcal E(\hat F_{\B}\otimes_g\hat A_{\tS}) \equiv \hat F_{\B}(\hat P_{\B})^{\deg \hat A_{\tS}}\otimes \hat A_{\tS}.
\end{align}
The extension mapping is isomorphic to the original one, that is 
\begin{align}\label{isomorphic}
  \mathcal E(\hat F_{\B}\otimes_g\hat A_{\tS})\mathcal E(\hat G_{\B}\otimes_g\hat B_{\tS}) = \mathcal E[(\hat F_{\B}\otimes_g\hat A_{\tS})(\hat G_{\B}\otimes_g\hat B_{\tS})].
\end{align}
The proof is presented in \App{AppB}. As a special case of \Eq{ext}, we see that for an arbitrary bath operator $\hat F_{\B}$, 
\begin{align}
  \mathcal E(\hat F_{\B}\otimes_g\mathbbm 1_{\tS}) = \hat F_{\B} \otimes \mathbbm 1_{\tS},
\end{align}
which is valid regardless of the parity of $\hat F_{\B}$. For an arbitrary system operator, 
\begin{align}
  \mathcal E(\mathbbm 1_{\B}\otimes_g\hat A_{\tS}) = \mathbbm 1_{\B}\otimes\hat A_{\tS}^{(+)} + \hat P_{\B}\otimes\hat A_{\tS}^{(-)},
\end{align}
with $\hat A_{\tS}^{(\pm)}$ being the even and odd parts of $\hat A_{\tS}$ defined by \Eq{eodec}. As a result, we denote $\hat F_{\B}$ representing a bath operator in both the pure-bath space $\mathcal H_{\B}$ and composite space $\mathcal H_{\T}$. And we denote 
\begin{align}
  \tilde A_{\tS} \equiv \hat A_{\tS}^{(+)} + \hat P_{\B}\otimes\hat A_{\tS}^{(-)}
\end{align}
to represent the system operator $\hat A_{\tS}$ in the total Hilbert space. Then we have 
\begin{align}
  H_{\SB} 
  &= \sum_\alpha\big( \hat F_\alpha\hat P_{\B}\otimes\hat s_\alpha^\dagger - \hat F_\alpha^\dagger\hat P_{\B}\otimes\hat s_\alpha \big),
\end{align}
by noting $\hat s_\alpha$ is of odd parity.

\subsection{Partial trace}

The open quantum system formalism focuses on the reduced system dynamics, which involves the tracing out of the bath degrees of freedom. We define the bath partial trace for a total space operator $\hat O_{\T}$ as
\begin{align}
  {\rm tr}_{\B}\hat O_{\T} = \sum_{\{n_{\alpha k}\}}\la \{n_{\alpha k}\}|\hat O_{\T}|\{n_{\alpha k}\}\ra.
\end{align}
Here, we use the bath occupation number basis, 
\begin{align}
  | \{n_{\alpha k}\} \ra \equiv \prod_{\alpha k} (c^{\dg}_{\alpha k})^{n_{\alpha k}} |\bf 0\ra.
\end{align}
Within this definition, we know that for any system and bath operators $\hat A_{\tS}$ and $\hat F_{\B}$, 
\begin{align}
  {\rm tr}_{\B}(\hat F_{\B}\otimes\hat A_{\tS}) = {\rm tr}_{\B}(\hat F_{\B}) \hat A_{\tS}.
\end{align}
More complicatedly, 
\begin{align}\label{AFB}
  {\rm Tr}(\tilde A_{\tS}\hat F_{\B}\otimes\hat B_{\tS}) &= {\rm tr}_{\B}(\hat F_{\B}){\rm tr}_{\tS}[\hat A_{\tS}\hat B_{\tS}^{(+)}] \nl 
  &\quad + {\rm tr}_{\B}(\hat P_{\B}\hat F_{\B}){\rm tr}_{\tS}[\hat A_{\tS}\hat B_{\tS}^{(-)}].
\end{align}
This can be proven by noting 
\begin{align}
  &\quad\,{\rm Tr}(\tilde A_{\tS}\hat F_{\B}\otimes\hat B_{\tS})\nl
  &= {\rm Tr}[\mathbbm 1_{\B}\otimes\hat A_{\tS}^{(+)}\hat F_{\B}\otimes\hat B_{\tS}] + {\rm Tr}[\hat P_{\B}\otimes\hat A_{\tS}^{(-)}\hat F_{\B}\otimes\hat B_{\tS}] \nl
  &= {\rm tr}_{\B}(\hat F_{\B}){\rm tr}_{\tS}[\hat A_{\tS}^{(+)}\hat B_{\tS}] + {\rm tr}_{\B}(\hat P_{\B}\hat F_{\B}){\rm tr}_{\tS}[\hat A_{\tS}^{(-)}\hat B_{\tS}]\nl
  &= {\rm tr}_{\B}(\hat F_{\B}){\rm tr}_{\tS}[\hat A_{\tS}\hat B_{\tS}^{(+)}] + {\rm tr}_{\B}(\hat P_{\B}\hat F_{\B}){\rm tr}_{\tS}[\hat A_{\tS}\hat B_{\tS}^{(-)}].
\end{align}
In the last identity, we have used the fact that 
\begin{align}
  {\rm tr}_{\tS}(\hat P_{\tS}^{\pm}\hat A_{\tS}\hat P_{\tS}^{\pm}\hat B_{\tS}) = {\rm tr}_{\tS}(\hat A_{\tS}\hat P_{\tS}^{\pm}\hat B_{\tS}\hat P_{\tS}^{\pm}),
\end{align}
with $\hat P_{\tS}^{\pm} \equiv \frac{1}{2}(1 \pm \hat P_{\tS})$; See the details in \App{AppA}. Using \Eq{AFB}, we obtain 
\begin{align}\label{AFB2}
  {\rm Tr}(\tilde{A}_{\tS}\hat F_{\B}\tilde B_{\tS}) = {\rm tr}_{\B}(\hat F_{\B}){\rm tr}_{\tS}(\hat A_{\tS}\hat B_{\tS}).
\end{align}

\subsection{Total space dynamics}

The system-plus-bath composite forms a closed quantum system, with the total density operator $\rho_{\T}(t)$ satisfying the Liouville-von Neumann equation
\begin{align}
  \dot\rho_{\T}(t) = -i[H_{\T},\rho_{\T}(t)].
\end{align}
The reduced system density operator is defined via 
\begin{align}\label{rhos}
  {\rm Tr}[\tilde A_{\tS}\rho_{\T}(t)] = {\rm tr}_{\tS}[\hat A_{\tS}\rho_{\tS}(t)]
\end{align}
for any system operator $\hat A_{\tS}$.  Quantum dissipation process always assumes the initial state as separate state, 
\begin{align}
  \rho_{\T}(0) = \mathcal E[\rho_{\B}^{\rm eq} \otimes_g \rho_{\tS}(0)] = \rho_{\B}^{\rm eq}\tilde\rho_{\tS}(0),
\end{align}
with 
\begin{align}
  \tilde\rho_{\tS}(0) = \rho_{\tS}^{(+)}(0) + \hat P_{\B}\rho_{\tS}^{(-)}(0)
\end{align}
and 
\begin{align}
  \rho_{\B}^{\rm eq} = \prod_{\alpha}\rho_{\alpha}^{\rm eq}, \quad \rho_\alpha^{\rm eq} = \frac{e^{-\beta_\alpha(h_{\alpha} - \mu_\alpha\hat N_\alpha)}}{{\rm tr}_{\B}e^{-\beta_\alpha(h_{\alpha} - \mu_\alpha\hat N_\alpha)}}.
\end{align}
being the grand canonical ensemble of the bath. Here, $\beta_\alpha = 1/(k_BT_{\alpha})$ and $\mu_\alpha$ ($\alpha = \rm L, R$) represents the inverse temperature and the chemical potential, respectively. From \Eq{AFB2}, we can verify the definition \Eq{rhos} holds for the separate state. However, it is worth noting that directly calculating the partial trace over the separate state gives 
\begin{align}
  \!\!{\rm tr}_{\B}\rho_{\T}(0) = \rho_{\tS}^{(+)}(0) + {\rm tr}_{\B}(\hat P_{\B}\rho_{\B}^{\rm eq})\rho_{\tS}^{(-)}(0) \neq \rho_{\tS}(0).
\end{align}
This indicates ${\rm tr}_{\B}\rho_{\T}(t)$ is incompatible to the definition of reduced system density operator [\Eq{rhos}], which is different to the bosonic scenario. \cite{Cir22035121}

\subsection{Reduced Dyson series}

To proceed, we turn to the interaction picture evolution, defined via 
\begin{align}\label{UI}
  U_I(t) \equiv e^{i(H_{\tS} + \tilde h_{\B})t}e^{-iH_{\T}t},
\end{align}
leading to 
\begin{align}
  \rho_{\T}(t) = e^{-i(H_{\tS} + \tilde h_{\B})t}\rho_{\T}^I(t) e^{i(H_{\tS} + \tilde h_{\B})t}
\end{align}
with 
\begin{align}\label{drhoI}
  \dot\rho_{\T}^I(t) = -i [H_{\SB}(t), \rho_{\T}^I(t)] 
\end{align}
and 
\begin{align}\label{HSBt}
  \!\!H_{\SB}(t) \equiv \sum_\alpha\big[\hat P_{\B}\hat F^\dagger_\alpha(t)\hat s_\alpha(t) - \hat P_{\B}\hat F_\alpha(t)\hat s_\alpha^\dagger(t)\big].
\end{align}
Here, $\hat F_\alpha(t) = e^{i\tilde h_{\B}t}\hat F_\alpha e^{-i\tilde h_Bt}$ and $\hat s_\alpha(t) = e^{iH_{\tS}t}\hat s_\alpha e^{-iH_{\tS}t}$. 
For convenience of notation, we denote the following superoperators for an operator $\hat O$, 
\begin{align}
  \begin{split}
    \hat O^>(\cdot) &\equiv \hat O(\cdot),\\
    \hat O^<(\cdot) &\equiv (\cdot)\hat O,\\
    \hat O^\times(\cdot) &\equiv [\hat O,(\cdot)] = (\hat O^> - \hat O^<)(\cdot).
  \end{split}
\end{align}
As an example, we can rewrite \Eq{drhoI} as
\begin{align}
  \dot\rho_{\T}^I(t) = -i H_{\SB}^{\times}(t)\rho_{\T}^I(t).
\end{align}
Its formal solution reads 
\begin{align}\label{rhoI}
  \rho_{\T}^I(t) &= \mathcal T\exp\bigg[ -i\int_{0}^t\!\!\ud\tau\,H_{\SB}^\times(\tau) \bigg]\rho_{\T}(0),
\end{align}
where we impose the time ordering for superoperators,
\begin{align}
  \mathcal T[H_{\SB}^\times(t_1)H_{\SB}^\times(t_2)] = 
  \begin{cases}
    H_{\SB}^\times(t_1)H_{\SB}^\times(t_2), &t_1\geq t_2,\\
    H_{\SB}^\times(t_2)H_{\SB}^\times(t_1), &t_1<t_2.
  \end{cases}
\end{align}
For deriving the reduced system density dynamics, our strategy goes by analyzing the structure of \Eq{rhoI} and expressing it in terms of sum of direct product of bath and system parts, 
\begin{align}\label{41}
  \rho_{\T}^I(t) = \sum_i \varrho_{\B}^i(t)\otimes \varrho_{\tS}^i(t).
\end{align}
From \Eq{rhos}, we have 
\begin{align}\label{rhos-def}
  &\quad\ \rho_{\tS}(t)\nl
  & = e^{-iH_{\tS}^\times t}\sum_i \Big\{ {\rm tr}_{\B}[\varrho_{\B}^i(t)]\varrho_{\tS}^{i(+)}(t) + {\rm tr}_{\B}[\hat P_{\B}\varrho_{\B}^i(t)]\varrho_{\tS}^{i(-)}(t) \Big\} \nl
  & \equiv e^{-iH_{\tS}^\times t}\rho_{\tS}^I(t),
\end{align} 
For simplifying the notation, we recast \Eq{HSBt} as 
\begin{align}
  H_{\SB}(t) = \sum_{\alpha\sigma}\sigma\hat P_{\B}\hat F_{\alpha}^\sigma(t)\hat s_\alpha^{\bar \sigma}(t).
\end{align}
Here, we introduce the notation $\sigma = \pm$, with $\hat F^{+}_\alpha = \hat F^\dagger_\alpha$ and $\hat F^-_\alpha = \hat F_\alpha$, and the system part being the same. We also denote $\bar\sigma\equiv -\sigma$.

Now we consider how the superoperator $H_{\SB}^\times(t)$ acts. Calculate 
\begin{align}
  H_{\SB}^\times(t) \hat O_{\B}\hat O_{\tS} &= \sum_{\alpha\sigma}\big[\sigma\hat P_{\B}\hat F_{\alpha}^\sigma(t)\hat O_{\B}\hat s_{\alpha}^{\bar \sigma}(t)\hat O_{\tS}\nl
  &\quad - \sigma\hat O_{\B}\hat P_{\B}\hat F_{\alpha}^\sigma(t)\hat O_{\tS}\hat s_{\alpha}^{\bar \sigma}(t)\big].
\end{align}
For $\hat O_{\B}\hat O_{\tS}$ being even, we have 
\begin{align}
  H_{\SB}^\times(t) \hat O_{\B}\hat O_{\tS} &= \sum_{\alpha\sigma}\big[\sigma\hat P_{\B}\hat F_{\alpha}^\sigma(t)\hat O_{\B}\hat s_{\alpha}^{\bar \sigma}(t)\hat O_{\tS}\nl
  &\quad - \sigma\hat P_{\B}\hat O_{\B}\hat P_{\B}\hat P_{\B}\hat F_{\alpha}^\sigma(t)\hat P_{\tS}\hat O_{\tS}\hat P_{\tS}\hat s_{\alpha}^{\bar \sigma}(t)\big].
\end{align}
Denote the following superoperators 
\begin{align}
  \!\!\mathscr B_{\alpha}^{\sigma>}(t)\hat O \equiv \hat P_{\B}\hat F_\alpha^\sigma(t)\hat O,\ \mathscr B_{\alpha}^{\sigma<}(t)\hat O \equiv \mathscr P_{\B}[\hat O\hat P_{\B}\hat F_\alpha^\sigma(t)]
\end{align}
and 
\begin{subequations}
  \begin{align}
  \mathscr s_{\alpha}^{\sigma>}(t)\hat O \equiv \bar\sigma\hat s_{\alpha}^{\sigma}(t)\hat O,\quad \mathscr s_{\alpha}^{\sigma<}(t)\hat O \equiv \sigma\mathscr P_{\tS}[\hat O\hat s_{\alpha}^{\sigma}(t)],\\
  \bar{\mathscr s}_{\alpha}^{\sigma>}(t)\hat O \equiv \bar\sigma\hat s_{\alpha}^{\sigma}(t)\hat O,\quad \bar{\mathscr s}_{\alpha}^{\sigma<}(t)\hat O \equiv \bar\sigma\mathscr P_{\tS}[\hat O\hat s_{\alpha}^{\sigma}(t)],
  \end{align}
\end{subequations}
with $\mathscr P_{\tS}(\cdot) \equiv \hat P_{\tS}(\cdot)\hat P_{\tS}$ and $\mathscr P_{\B}(\cdot) \equiv \hat P_{\B}(\cdot)\hat P_{\B}$. For later use, we use the index $\lambda$ to label the left and right action, that is $\lambda = +1$ for $>$ and $\lambda = -1$ for $<$, and $\bar\lambda = -\lambda$. Then we have 
\begin{align}
  H_{\SB}^\times(t) \hat O_{\B}\hat O_{\tS} = \sum_{\alpha\sigma\lambda}\mathscr B_{\alpha}^{\sigma\lambda}(t)\hat O_{\B}\mathscr s_{\alpha}^{\bar\sigma\lambda}(t)\hat O_{\tS},
\end{align}
for $\hat O_{\B}\hat O_{\tS}$ even, and 
\begin{align}
  H_{\SB}^\times(t) \hat O_{\B}\hat O_{\tS} = \sum_{\alpha\sigma\lambda}\mathscr B_{\alpha}^{\sigma\lambda}(t)\hat O_{\B}\bar{\mathscr s}_{\alpha}^{\bar\sigma\lambda}(t)\hat O_{\tS},
\end{align}
for $\hat O_{\B}\hat O_{\tS}$ odd. As a result, the first order contribution reads
\begin{align}\label{50}
  H_{\SB}^\times(t)\rho_{\T}(0) &= \sum_{\alpha\sigma\lambda} \mathscr B_{\alpha}^{\sigma\lambda}(t)\rho_{\B}^{\rm eq}\mathscr s_{\alpha}^{\bar\sigma\lambda}(t) \rho_{\tS}^{(+)}(0) \nl
  &\quad + \sum_{\alpha\sigma\lambda} \mathscr B_{\alpha}^{\sigma\lambda}(t)(\rho_{\B}^{\rm eq}\hat P_{\B})\bar{\mathscr s}_{\alpha}^{\bar\sigma\lambda}(t) \rho_{\tS}^{(-)}(0).
\end{align}
Since each action of $H^\times_{\SB}(t)$ remains the parity of operators, we can rewrite \Eq{rhoI} as 
\begin{widetext}
\begin{align}\label{51}
  \rho_{\T}^I(t) = \sum_{n=0}^\infty\frac{(-i)^n}{n!}&\int_{0}^t\prod_{i=1}^n\ud t_i\sum_{\alpha_1\sigma_1\lambda_1,\cdots,\alpha_n\sigma_n\lambda_n} \bigg\{ \big[\mathcal T_{\B}\mathscr B_{\alpha_n}^{\sigma_n\lambda_n}(t_n) \cdots \mathscr B_{\alpha_1}^{\sigma_1\lambda_1}(t_1)\rho_{\B}^{\rm eq}\big]\big[\mathcal T_{\tS}\mathscr s_{\alpha_n}^{\bar\sigma_n\lambda_n}(t_n)\cdots \mathscr s_{\alpha_1}^{\bar\sigma_1\lambda_1}(t_1) \rho_{\tS}^{(+)}(0) \big] \nl
  &\quad +  \big[\mathcal T_{\B}\mathscr B_{\alpha_n}^{\sigma_n\lambda_n}(t_n) \cdots \mathscr B_{\alpha_1}^{\sigma_1\lambda_1}(t_1)(\rho_{\B}^{\rm eq}\hat P_{\B})\big]\big[\mathcal T_{\tS}\bar{\mathscr s}_{\alpha_n}^{\bar\sigma_n\lambda_n}(t_n)\cdots \bar{\mathscr s}_{\alpha_1}^{\bar\sigma_1\lambda_1}(t_1) \rho_{\tS}^{(-)}(0) \big]  \bigg\}.
\end{align}
Here, we separate the total time ordering to $\mathcal T = \mathcal T_{\B}\mathcal T_{\tS}$, with the system and bath time ordering being fermionic type, i.e., 
\begin{align}
  \mathcal T_{\B}[\mathscr B_{\alpha_2}^{\sigma_2\lambda_2}(t_2)\mathscr B_{\alpha_1}^{\sigma_1\lambda_1}(t_1)] = 
  \begin{cases}
    \mathscr B_{\alpha_2}^{\sigma_2\lambda_2}(t_2)\mathscr B_{\alpha_1}^{\sigma_1\lambda_1}(t_1), &\quad t_2\geq t_1, \\ 
    -\mathscr B_{\alpha_1}^{\sigma_1\lambda_1}(t_1)\mathscr B_{\alpha_2}^{\sigma_2\lambda_2}(t_2), &\quad t_2<t_1,
  \end{cases}
\end{align}
and the system one is similar. By comparing \Eq{51} with \Eq{41} and using \Eq{rhos-def}, we readily have 
\begin{align}\label{53}
  \rho_{\tS}^I(t) = \sum_{n=0}^\infty\frac{(-i)^n}{n!}\int\prod_{i=1}^n\ud t_i&\sum_{\alpha_1\sigma_1\lambda_1,\cdots,\alpha_n\sigma_n\lambda_n}\bigg\{ {\rm tr}_{\B}\big[\mathcal T_{\B} \mathscr B_{\alpha_n}^{\sigma_n\lambda_n}(t_n) \cdots \mathscr B_{\alpha_1}^{\sigma_1\lambda_1}(t_1)\rho_{\B}^{\rm eq} \big]\big[\mathcal T_{\tS}\mathscr s_{\alpha_n}^{\bar\sigma_n\lambda_n}(t_n)\cdots \mathscr s_{\alpha_1}^{\bar\sigma_1\lambda_1}(t_1) \rho_{\tS}^{(+)}(0) \big] \nl
  &\quad + {\rm tr}_{\B}\big[\mathcal T_{\B} \hat P_{\B}\mathscr B_{\alpha_n}^{\sigma_n\lambda_n}(t_n) \cdots \mathscr B_{\alpha_1}^{\sigma_1\lambda_1}(t_1)(\rho_{\B}^{\rm eq}\hat P_{\B}) \big]\big[\mathcal T_{\tS}\mathscr s_{\alpha_n}^{\bar\sigma_n\lambda_n}(t_n)\cdots \mathscr s_{\alpha_1}^{\bar\sigma_1\lambda_1}(t_1) \rho_{\tS}^{(-)}(0) \big] \bigg\}.
\end{align}
For further proceeding, we have to consider the multi-point correlation functions ${\rm tr}_{\B}\big[ \mathcal T_{\B}\mathscr B_{\alpha_n}^{\sigma_n\lambda_n}(t_n) \cdots \mathscr B_{\alpha_1}^{\sigma_1\lambda_1}(t_1)\rho_{\B}^{\rm eq} \big]$ and ${\rm tr}_{\B}\big[\mathcal T_{\B} \hat P_{\B}\mathscr B_{\alpha_n}^{\sigma_n\lambda_n}(t_n) \cdots \mathscr B_{\alpha_1}^{\sigma_1\lambda_1}(t_1)(\rho_{\B}^{\rm eq}\hat P_{\B}) \big]$. Note that for $n$ being odd, the correlation functions vanish. Thus, we only focus on the even $n$ case. Based on $\{\hat P_{\B},\hat F_\alpha^\sigma(t)\} = 0$ and $\hat P_{\B}^2 = 1$, those even $n$-point correlations reduce to ${\rm tr}_{\B}\big[\mathcal T_{\B} \mathscr F_{\alpha_n}^{\sigma_n\lambda_n}(t_n) \cdots \mathscr F_{\alpha_1}^{\sigma_1\lambda_1}(t_1)\rho_{\B}^{\rm eq} \big]$ and ${\rm tr}_{\B}\big[ \mathcal T_{\B}\hat P_{\B}\mathscr F_{\alpha_n}^{\sigma_n\lambda_n}(t_n) \cdots \mathscr F_{\alpha_1}^{\sigma_1\lambda_1}(t_1)(\rho_{\B}^{\rm eq}\hat P_{\B}) \big]$, 
with 
\begin{align}
  \mathscr F^{\sigma>}_\alpha(t)\hat O \equiv \hat F_{\alpha}(t)\hat O,\quad \mathscr F^{\sigma<}_\alpha(t)\hat O \equiv \mathscr P_{\B}[\hat O\hat F_{\alpha}(t)].
\end{align}
In order to combine the two large terms in the summation of \Eq{53}, we observe that the system part of the second term differs from the first one only by minus signs of number of the right actions $\bar{\mathscr{s}}_\alpha^{\sigma <}(t)$. On the other hand, in order to take the $\hat P_{\B}$ within ${\rm tr}_{\B}\big[\mathcal T_{\B} \hat P_{\B}\mathscr F_{\alpha_n}^{\sigma_n\lambda_n}(t_n) \cdots \mathscr F_{\alpha_1}^{\sigma_1\lambda_1}(t_1)(\rho_{\B}^{\rm eq}\hat P_{\B}) \big]$, we have to compensate the minus signs of number of the left actions $\mathscr F^{\sigma>}_{\alpha}(t)$. Since the total contribution is equal to $(-1)^n$ with $n$ being even (only even $n$ is considered), then no additional sign is introduced. As a result, we have 
\begin{align}\label{eq55}
  \rho_{\tS}^I(t) &= \sum_{n=0}^\infty\frac{(-i)^n}{n!}\int\prod_{i=1}^n\ud t_i\sum_{\alpha_1\sigma_1\lambda_1,\cdots,\alpha_n\sigma_n\lambda_n} {\rm tr}_{\B}\big[ \mathcal T_{\B}\mathscr F_{\alpha_n}^{\sigma_n\lambda_n}(t_n) \cdots \mathscr F_{\alpha_1}^{\sigma_1\lambda_1}(t_1)\rho_{\B}^{\rm eq} \big]\big[\mathcal T_{\tS}\mathscr s_{\alpha_n}^{\bar\sigma_n\lambda_n}(t_n)\cdots \mathscr s_{\alpha_1}^{\bar\sigma_1\lambda_1}(t_1) \rho_{\tS}^{}(0) \big]\nl
  &\equiv \sum_{n=0}^\infty\frac{(-i)^n}{n!}\int\prod_{i=1}^n\ud t_i\sum_{\alpha_1\sigma_1\lambda_1,\cdots,\alpha_n\sigma_n\lambda_n}C_{\alpha_n,\cdots,\alpha_1}^{\sigma_n\lambda_n,\cdots,\sigma_1\lambda_1}(t_n,\cdots,t_1)\big[\mathcal T_{\tS}\mathscr s_{\alpha_n}^{\bar\sigma_n\lambda_n}(t_n)\cdots \mathscr s_{\alpha_1}^{\bar\sigma_1\lambda_1}(t_1) \rho_{\tS}^{}(0) \big]
\end{align}
with noting $\rho_{\tS}(0) = \rho_{\tS}^{(+)}(0) + \rho_{\tS}^{(-)}(0)$. So far, we derive the reduced system density operator in terms of the Dyson series. Nextly, we apply the Wick's theorem for fermionic superoperators to resolve the multi-point correlation functions.

\end{widetext}

\subsection{Wick's theorem and influence functional}

The Wick's theorem for fermionic superoperators is reviewed in \App{AppB}. It states that the multi-point correlation function can be expressed as the sum of all possible products of two-point ones. As a result, for even $n$, we have
\begin{align}\label{eq56}
  &\quad \ C_{\alpha_n,\cdots,\alpha_1}^{\sigma_n\lambda_n,\cdots,\sigma_1\lambda_1}(t_n,\cdots,t_1)\nl
  &= \sum_{\mathscr c\in C_n}(-)^{\#_\mathscr c}\prod_{(i, j)\in\mathscr c}C_{\alpha_i\alpha_j}^{\sigma_i\lambda_i,\sigma_j\lambda_j}(t_i,t_j),
\end{align}
with 
\begin{align}
  C_{\alpha_2\alpha_1}^{\sigma_2\lambda_2,\sigma_1\lambda_1}(t_2,t_1) = {\rm tr}_{\B}\big[ \mathcal T_{\B} \mathscr F_{\alpha_2}^{\sigma_2\lambda_2}(t_2) \mathscr F_{\alpha_1}^{\sigma_1\lambda_1}(t_1)\rho_{\B}^{\rm eq} \big].
\end{align}
Here, $C_n$ is the set of all possible time-ordered pairings and $\#_\mathscr c$ counts the crossing number of the pairing configuration $\mathscr c$. Thus, the reduced system density operator is recast as
\begin{align}\label{58}
  \rho_{\tS}^I(t) = \sum_{n=0}^\infty\frac{(-i)^{2n}}{(2n)!}\!\!\sum_{\mathscr c\in C_{2n}}\prod_{(i, j)\in\mathscr c}\!\!\mathcal T_{\tS}\int_{0}^{t}\!\!\ud t_2\ud t_1\,\mathcal W(t_2,t_1)\rho_{\tS}(0),
\end{align}
where 
\begin{align}\label{W21}
  \mathcal W(t_2,t_1) &\equiv \!\!\!\!\sum_{\alpha_1\sigma_1\lambda_1,\alpha_2\sigma_2\lambda_2}\!\!\!\!\!C_{\alpha_2\alpha_1}^{\sigma_2\lambda_2,\sigma_1\lambda_1}(t_2,t_1)\mathscr s_{\alpha_2}^{\bar\sigma_2\lambda_2}(t_2)\mathscr s_{\alpha_1}^{\bar\sigma_1\lambda_1}(t_1)\nl
  &= -\sum_{\alpha\sigma}\mathscr A^{\bar\sigma}_{\alpha}(t_2)\mathscr C^{\sigma}_{\alpha}(t_2, t_1),
\end{align}
with 
\begin{align}
  \mathscr A^{\sigma}_{\alpha}(t)\hat O &\equiv \hat s_{\alpha}^\sigma(t)\hat O - \mathscr P_{\tS}[\hat O\hat s_{\alpha}^\sigma(t)], \\
  \mathscr C^{\sigma}_{\alpha}(t_2, t_1)\hat O &\equiv C^\sigma_\alpha(t_2 - t_1)\hat s^\sigma_\alpha(t_1)\hat O\nl
  &\quad\  + C_\alpha^{\bar\sigma*}(t_2 - t_1)\mathscr P_{\tS}[\hat O\hat s_{\alpha}^\sigma(t_1)].
\end{align}
and the bare-bath correlation function being
\begin{align}
  C_{\alpha}^\sigma(t) \equiv {\rm tr}_{\B}[\hat F_{\alpha}^\sigma(t)\hat F_{\alpha}^{\bar\sigma}(0)\rho_{\B}^{\rm eq}] \equiv \la\hat F_{\alpha}^\sigma(t)\hat F_{\alpha}^{\bar\sigma}(0)\ra_{\B}.
\end{align}
Note that the minus signs from $\mathscr c$ are exactly compensated by the fermionic time ordering of system superoperators. Since all terms in the product of \Eq{58} are identical and the total number of those terms is $(2n-1)!!$, then we have 
\begin{align}\label{influence_functional}
  \rho_{\tS}^I(t) &= \sum_{n=0}^\infty\frac{(-i)^{2n}}{(2n)!}(2n-1)!!\mathcal T_{\tS}2^n\mathcal F^n\rho_{\tS}(0)\nl
  &= \mathcal T_{\tS}e^{\mathcal F}\rho_{\tS}(0),
\end{align}
with the influence superoperator being
\begin{align}
  \mathcal F \equiv \int_{0}^t\!\ud t_2\int_0^{t_2}\!\ud t_1\,\mathcal W(t_2,t_1).
\end{align}
In the last identity of \Eq{influence_functional}, we have used the fact that $(2n)!/(2^n n!) = (2n-1)!!$.

\section{Hierarchical equations of motion}\label{sec4}


\subsection{Exponential decomposition of bath correlation functions}

Quantum dissipation dynamics aims at constructing the dynamics of reduced system density operator. Based on the influence functional formalism, one intuitive way is to differentiate \Eq{influence_functional}. However, such operations generally lead to more auxiliary quantities which are not closed. The key to resolve this issue is to expand the bath correlation functions in exponential series for $t>0$, 
\begin{align}
  C_{\alpha}^{\sigma}(t) = \sum_{\kappa = 1}^K \eta_{\alpha\kappa}^{\sigma} e^{-\gamma_{\alpha\kappa}^{\sigma} t},
\end{align}
with the prefactor $\eta_{\alpha\kappa}^{\sigma}$ and exponential $\gamma_{\alpha\kappa}^{\sigma}$ being complex. The exponential decomposition can be achieved via various schemes, such as the Matsubara decomposition,\cite{Tan906676,Xu05041103} the Pad\'e spectrum decomposition,\cite{Hu10101106,Hu11244106} time-domain Prony scheme,\cite{Che22221102} the numerically analytical continuation,\cite{Zha25214111} and so on.\cite{Cui19024110} All the decomposition schemes require \cite{Jin08234703,Yan16110306}
\begin{align}
  \gamma_{\alpha\kappa}^{\sigma*} = \gamma_{\alpha\kappa}^{\bar\sigma}.
\end{align}
This leads to
\begin{align}
  C_{\alpha}^{\bar\sigma*}(t) = \sum_{\kappa}\eta_{\alpha\kappa}^{\bar\sigma*}e^{-\gamma_{\alpha\kappa}^{\bar\sigma*}t} = \sum_{\kappa}\eta_{\alpha\kappa}^{\bar\sigma*}e^{-\gamma_{\alpha\kappa}^{\sigma}t}.
\end{align}
In practical, we determine the bath correlation functions via the fluctuation--dissipation theorem,
\begin{align}
  C_{\alpha}^\sigma(t) = \frac{1}{\pi}e^{i\sigma\varphi_\alpha t}\int_{-\infty}^\infty\!\!\ud\omega\,e^{i\sigma\omega t}\frac{\Gamma_{\alpha}^\sigma(\omega)}{1 + e^{\sigma\beta(\omega - \mu_\alpha)}},
\end{align}
where the spectral density function is defined via 
\begin{align}
  \Gamma_\alpha^\sigma(\omega) = \pi\sum_k |t_{\alpha k}|^2\delta(\omega - \epsilon_k).
\end{align}
As a result, we recast \Eq{W21} as
\begin{align}
  \mathcal W(t_2, t_1) = -\sum_{\alpha\sigma} \mathscr A_{\alpha}^{\bar\sigma}(t_2)\sum_{\kappa}e^{-\gamma_{\alpha\kappa}^\sigma(t_2 - t_1)}\mathscr C_{\alpha\kappa}^{\sigma}(t_1)
\end{align}
with 
\begin{align}
  \mathscr C^{\sigma}_{\alpha\kappa}(t) \hat O \equiv \eta_{\alpha\kappa}^\sigma \hat s_{\alpha}^\sigma(t)\hat O + \eta_{\alpha\kappa}^{\bar\sigma*}\mathscr P_{\tS}[\hat O\hat s_{\alpha}^\sigma(t)].
\end{align}

\subsection{Construction of HEOM}

We are now in the position to construct the hierarchical equations of motion (HEOM). We introduce the index abbreviation 
\begin{align}
  j \equiv (\sigma\alpha\kappa),\quad \bar j \equiv (\bar\sigma\alpha\kappa).
\end{align}
Define the auxiliary density operators (ADOs) as
\begin{align}
  \rho_{\bf j}^{(n)}(t) \equiv e^{-iH_{\tS}^\times t}\mathcal T_{\tS}\mathcal D_{j_n}\cdots\mathcal D_{j_1}e^{\mathcal F}\rho_{\tS}(0),
\end{align}
where 
\begin{align}
  \mathcal D_{j} \equiv \mathcal D_{\alpha\kappa}^\sigma \equiv -i \int_0^t\!\!\ud\tau\, e^{-\gamma_{\alpha\kappa}^\sigma(t - \tau)}\mathscr C_{\alpha\kappa}^\sigma(\tau)
\end{align}
and $\mathbf j \equiv j_1j_2\cdots j_n$. Using 
\begin{align}
  \frac{\ud}{\ud t}\mathcal T_{\tS}e^{\mathcal F} = -i \sum_{\alpha\sigma\kappa}\mathscr A_{\alpha}^{\bar\sigma}(t)\mathcal D_{\alpha\kappa}^\sigma e^{\mathcal F}\rho_{\tS}(0),
\end{align}
we have 
\begin{align}
  \dot\rho_{\bf j}^{(n)} &= - \bigg( iH_{\tS}^\times + \sum_{r=1}^n\gamma_{j_r} \bigg)\rho_{\bf j}^{(n)} - i \sum_{j}\mathcal A_{\bar j}\rho_{\mathbf jj}^{(n+1)} \nl
  &\quad\, -i \sum_{r=1}^n (-)^{n-r} \mathcal C_{j_r}\rho_{{\bf j}_r^-}^{(n-1)}. 
\end{align}
with $\mathbf j_r^-$ being the index string by removing $j_r$ from $\mathbf j$. The reduced system density operator is just the zeroth-tier ADO, i.e., $\rho_{\tS}(t) = \rho^{(0)}(t)$. The initial conditions are $\rho^{(0)}(0) = \rho_{\tS}(0)$ and $\rho_{}^{(n>0)}(0) = 0$. Here, the superoperators are defined as 
\begin{subequations}
  \begin{align}
  \mathcal A_{j}\hat O &\equiv \hat s_{\alpha}^\sigma\hat O - \mathscr P_{\tS}(\hat O\hat s_{\alpha}^\sigma), \\
  \mathcal C_{j}\hat O &\equiv \eta_{\alpha\kappa}^\sigma\hat s_{\alpha}^\sigma\hat O + \eta_{\alpha\kappa}^{\bar\sigma*}\mathscr P_{\tS}(\hat O\hat s_{\alpha}^\sigma).
  \end{align}
\end{subequations}
Generally, since the initial reduced system density operator is physical, it must be an even one. In such case, the superoperators are given by
\begin{subequations}
  \begin{align}
    \mathcal A_{j} \rho^{(n\pm1)} &= \hat s_{\alpha}^\sigma\rho^{(n\pm1)} - (-)^n\rho^{(n\pm1)}\hat s_{\alpha}^\sigma, \\
    \mathcal C_{j} \rho^{(n\pm1)} &\equiv \eta_{\alpha\kappa}^\sigma\hat s_{\alpha}^\sigma\rho^{(n\pm1)} + (-)^{n}\eta_{\alpha\kappa}^{\bar\sigma*}\rho^{(n\pm1)}\hat s_{\alpha}^\sigma.
  \end{align}
\end{subequations}
\subsection{Transport current}

\begin{figure*}[ht]
  \centering
  \includegraphics[width=0.72\textwidth]{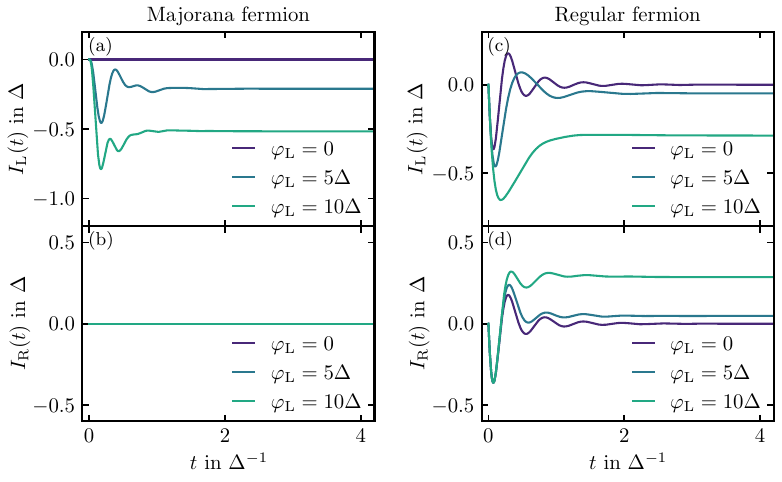}
  \caption{Transient dynamics of transport current for Majorana impurity [(a) and (b)] and regular fermion [(c) and (d)]. We set $\varepsilon_{\rm  M} = \varepsilon_{\rm F} = 10\Delta$ and $\beta_{\rm L} = \beta_{\rm R} = 1000\Delta^{-1}$. The values of external bias voltages are given as $\varphi_{\rm L} = 0, 5\Delta, 10\Delta$ and $\varphi_{\rm R} = 0$, respectively. Remarkably, the Majorana impurity model exhibits a non-vanishing steady-state current when $\varphi_{\rm L}$ is absent. And the FBO model does not have such a feature, since its total particle number is conserved at the steady-state.}
  \label{fig2}
\end{figure*}

For discussing the transport property of the impurity system, we have to consider the dynamics of electronic current, defined as \cite{Mei922512,Jin22093009,Yan16110306}
\begin{align}
  \!\hat I_\alpha \equiv -\frac{\ud \hat N_{\alpha}}{\ud t} = i(\tilde s_\alpha^\dagger\hat F_\alpha - \hat F_\alpha^\dagger\tilde s_\alpha) = i \sum_\sigma \hat s_\alpha^{\bar\sigma}\otimes\hat P_{\B}\hat F_{\alpha}^\sigma,
\end{align}
with $\hat N_{\alpha} = \sum_k\hat c_{\alpha k}^\dagger\hat c_{\alpha k}$ being the particle number of the $\alpha$-th bath. We are interested in the transient mean value of the current, 
\begin{align}
  I_\alpha(t) \equiv {\rm Tr}[\hat I_{\alpha}\rho_{\T}(t)] = i\sum_\sigma{\rm Tr}[\hat s^{\bar\sigma}_\alpha\hat P_{\B}\hat F^\sigma_\alpha\rho_{\T}(t)].
\end{align}
By defining the interaction picture \Eq{UI}, we have 
\begin{align}
  I_{\alpha}(t) = i\sum_\sigma{\rm tr}_{\tS}[\hat s_{\alpha}^{\bar\sigma}(t)\varrho_{\tS}^{\sigma\alpha}(t)],
\end{align}
where $\varrho_{\tS}^{\sigma\alpha}(t)$ is the obtained via tracing out the bath degrees of freedom of 
\begin{align}
  \varrho_{\T}^{\sigma\alpha}(t) \equiv \hat P_{\B}\hat F^\sigma_{\alpha}(t)\rho_{\T}^I(t) = \mathscr B^{\sigma >}_\alpha(t)\rho_{\T}^I(t).
\end{align}
Using \Eq{50}, we have 
\begin{align}
  \varrho_{\T}^{\sigma\alpha}(t) = i\frac{\delta}{\delta\mathscr s^{\bar\sigma>}_\alpha(t)}\rho_{\T}^{I}(t),
\end{align}
where ${\delta}/{\delta\mathscr s^{\bar\sigma>}_\alpha(t)}$ is the functional derivative over the superoperator $\mathscr s^{\bar\sigma>}_\alpha(t)$. Consequently, we have 
\begin{align}\label{85}
  \varrho_{\tS}^{\sigma\alpha}(t) = i\frac{\delta}{\delta\mathscr s^{\bar\sigma>}_\alpha(t)}\rho_{\tS}^I(t) = i\sigma\frac{\delta}{\delta\hat s^{\bar\sigma>}_{\alpha}(t)}\rho_{\tS}^I(t).
\end{align}
Substituting \Eq{influence_functional} into \Eq{85}, one readily obtain 
\begin{align}
  \varrho_{\tS}^{\sigma\alpha}(t) = -\sigma\sum_{\kappa=1}^K \mathcal T_{\tS}\mathcal D_{\alpha \kappa}^\sigma e^{\mathcal F}\rho_{\tS}(0).
\end{align}
Then, the current is evaluated by using the first tier of the ADOs, namely, 
\begin{align}
  I_{\alpha}(t) = -i\sum_{\sigma}\sum_{\kappa}{\rm tr}_{\tS}[\sigma\hat s^{\bar\sigma}_{\alpha}\rho_{\sigma\alpha\kappa}^{(1)}(t)].
\end{align}
Thus, we have finished the HEOM formalism for evaluating the transport current from the canonical algebra.

\section{Numerical demonstration}\label{sec5}

In this section, we apply the HEOM method to demonstrate the transport phenomena induced by a Majorana impurity. For both two baths, we adopt the spectral density function being the Lorentz type, namely
\begin{align}
  \Gamma^\sigma_\alpha(\omega) = \frac{\Delta W^2}{\omega^2 + W^2}.
\end{align}
We set the chemical potential as the zero energy point, $\mu_{\rm L} = \mu_{\rm R} = 0$. The band width is set to be $W = 10\Delta$. For illustrating the unique Majorana transport property, we will compare the results with a usual fermionic Brownian oscillator (FBO) model, 
\begin{align}
  H_{\rm FBO} = \varepsilon_{\rm F}\hat a^\dagger\hat a + \tilde h_{\B} + \sum_{\alpha}(\hat a\hat F^\dagger_{\alpha} + \hat F_{\alpha}\hat a^\dagger)
\end{align}
with $\hat a$ ($\hat a^\dagger$) being the annihilation (creation) operator of the impurity fermion and $\varepsilon_{\rm F}$ being its energy level.

In \Fig{fig2}, we present the transient current dynamics. The parameters are set to $\varepsilon_{\rm M} = \varepsilon_{\rm F} = 10\Delta$ and $\beta_{\rm L} = \beta_{\rm R} = 1000\Delta^{-1}$. The external bias voltages are chosen as $\varphi_{\rm L} = 0, 5\Delta, 10\Delta$ while $\varphi_{\rm R} = 0$. In both models, the transient currents exhibit oscillations at frequency around the system eigenenergy, and the magnitude of the steady-state current $|I_{\rm L}^{\rm st}|$ increases with $\varphi_{\rm L}$. In contrast, the Majorana impurity displays behavior distinct from the FBO model: (i) in the absence of an applied bias, the transient current vanishes for the Majorana impurity, whereas the FBO model still shows a finite current; (ii) the steady-state total current, $I_{\rm L}^{\rm st}+I_{\rm R}^{\rm st}$, is nonzero for the Majorana case but vanishes for the FBO at long times. This originates from the fact that, for the Majorana impurity, the total current is not a conserved quantity of the total Hamiltonian.

\begin{figure}[t]
  \centering
  \includegraphics[width=0.4\textwidth]{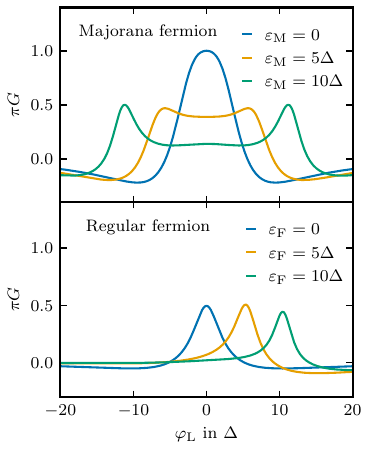}
  \caption{The steady-state differential conductance as a function of bias voltage at various values ($0,5\Delta,10\Delta$) of system energy level. The parameters are set to $\beta_{\rm L} = \beta_{\rm R} = 1000\Delta^{-1}$ and $\varphi_{\rm R} = 0$.}
  \label{fig3}
\end{figure}

Figure \ref{fig3} presents the steady-state differential conductance, $G \equiv -{\ud I_{\rm L}^{\rm st}}/{\ud \varphi_{\rm L}}$, at various values of system energy level. The FBO model exhibits a single conductance peak located at the system energy level, $\varepsilon_{\rm F}$. By contrast, the Majorana model shows more intricate behavior: as $\varepsilon_{\rm M}$ increases, the conductance peak splits into two, with positions approximately at $\pm\varepsilon_{\rm M}$. Notably, for the non-interacting zero mode ($\varepsilon_{\rm M}=0$), the peak height agrees with the Landauer--B\"uttiker result, $G(\varphi_{\rm L})=1/\pi$, a hallmark of the Majorana zero mode.
\cite{Liu11201308}

\section{Summary}\label{sec6}

In summary, we have established the HEOM formalism for quantum transport through a Majorana impurity system. The present theory is constructed based on the canonical algebra and the fermionic superoperators. The key step is to introduce the Wick's theorem for fermionic superoperators. Based on this, we are allowed to sum over the contributions of all the bath degrees of freedom and obtain the influence functional. The HEOM formalism is then constructed via the exponential decomposition of bath correlation functions. Besides, we use the functional derivative technique to construct the relation between the transport current and the first-tier ADOs. Generalizing the technique would lead to the influence functional representation of the generalized Wick's theorem for fermionic dissipatons.\cite{Yan14054105,Yan16110306,Su23024113,Su254107}

For numerical demonstration, we have investigated the transient transport dynamics and steady-state differential conductance of a Majorana impurity system. The results are compared with those of a regular fermionic impurity model. The unique transport properties induced by the Majorana mode are clearly exhibited. The present HEOM formalism is numerically exact and applicable to arbitrary system--bath coupling strength, external bias voltage, and temperature. It provides a powerful tool to explore the exotic transport phenomena in Majorana systems.

\begin{acknowledgments}
Support from the National Natural Science Foundation of China (Grant Nos.\ 224B2305, 22373091), and the Innovation Program for Quantum Science and Technology (Grant No.\ 2021ZD0303301) is gratefully acknowledged.
The authors are indebted to Prof.\,YiJing Yan for invaluable discussions.
\end{acknowledgments}

\appendix

\section{Structure of fermionic Hilbert space}\label{AppA}

This appendix discusses the algebraic structure of a fermionic Hilbert space. The $N$-fermion Hilbert space is spanned by the occupation number basis, 
\begin{align}
  |n_1,n_2,\cdots, n_N\ra \equiv \hat c_{1}^{\dagger n_1}\hat c_{2}^{\dagger n_2}\cdots\hat c_{N}^{\dagger n_N}|{\bf 0}\ra.
\end{align}
Here, we introduce the no-particle vacuum state $|\bf 0\ra$ and the creation and annihilation operators with anti-commutation relations,
\begin{align}
  \{\hat c_k, \hat c_{k'}^\dagger\} = \delta_{kk'}, \quad \{\hat c_k, \hat c_{k'}\} = \{\hat c_{k}^\dagger, \hat c_{k'}^\dagger\} = 0.
\end{align}
The fermionic statistics requires the occupation number for each state can only be 0 or 1. We can separate the total Hilbert space into two parts via determining thektotal particle number $n \equiv \sum_k n_k$ is odd or even. To achieve this, we define the parity operator
\begin{align}
  \hat P \equiv \prod_k \exp(i\pi\hat c_k^\dagger\hat c_k),
\end{align}
leading to 
\begin{align}
  \!\!\!\!\!\hat P|n_1,\cdots,n_N\ra\! =\! 
  \begin{cases}
    |n_1,\cdots,n_N\ra, &\text{$\sum_i n_i$ is even}, \\
    -|n_1,\cdots,n_N\ra, &\text{$\sum_i n_i$ is odd}.
  \end{cases}
\end{align}
We can always project a quantum state into the odd and even part by introducing the projection operators
\begin{align}
  \hat P_{+} \equiv \frac{1}{2}(1 + \hat P)\quad\text{and}\quad\hat P_{-} \equiv \frac{1}{2}(1 - \hat P),
\end{align}
with noting $\hat P_{\pm}^2 = \hat P_{\pm}$, $\hat P_+\hat P_- = 0$, and $\hat P_+ + \hat P_- = \hat I$. 

Consequently, an operator $\hat O$ is also decomposed into the even and odd parts,  
\begin{align}\label{eodec}
  \hat O = \hat O^{(+)} + \hat O^{(-)},
\end{align}
with 
\begin{align}\label{defeo}
  \begin{split}
    \hat O^{(+)} &\equiv \hat P_+\hat O\hat P_+ + \hat P_-\hat O\hat P_- = \frac{1}{2}(\hat O + \hat P\hat O\hat P), \\
    \hat O^{(-)} &\equiv \hat P_+\hat O\hat P_- + \hat P_-\hat O\hat P_+ = \frac{1}{2}(\hat O - \hat P\hat O\hat P).
  \end{split}
\end{align}
Evidently, ${\rm Tr}\,\hat O^{(+)} = {\rm Tr}\,\hat O$ and ${\rm Tr}\,\hat O^{(-)} = 0$. It is easy to see from \Eq{defeo} that an operator $\hat O$ is an even one if and only if $[\hat O,\hat P] = 0$; conversely, $\hat O$ is odd if and only if $\{\hat O,\hat P\} = 0$. As a corollary, the product of two even or two odd operators is an even one; the product of an even and an odd operators is odd. As an example, $\hat c_k$ is odd and $\hat c_k^\dagger\hat c_k$ is even.  

\section{Proof of \Eq{isomorphic}} \label{AppB}

We proof it directly. The left-hand side of \Eq{isomorphic} is
\begin{align}
  \mathcal E(\hat F_{\B}\!\otimes_g\!\hat A_{\tS})\mathcal E(\hat G_{\B}\!\otimes_g\!\hat B_{\tS}) = \hat F_{\B}\hat P_{\B}^{\deg\hat A_{\tS}}\hat G_{\B}\hat P_{\B}^{\deg\hat B_{\tS}}\!\otimes\! \hat A_{\tS}\hat B_{\tS}.
\end{align}
The right-hand side is
\begin{align}
  &\quad \ \mathcal E\big[(\hat F_{\B}\!\otimes_g\!\hat A_{\tS})(\hat G_{\B}\!\otimes_g\!\hat B_{\tS})\big] \nl
  &= \mathcal E\big[(-1)^{\deg\hat A_{\tS}\deg\hat G_{\B}}\hat F_{\B}\hat G_{\B}\!\otimes_g\! \hat A_{\tS}\hat B_{\tS}\big] \nl
  &= (-1)^{\deg\hat A_{\tS}\deg\hat G_{\B}}\hat F_{\B}\hat G_{\B}\hat P_{\B}^{\deg(\hat A_{\tS}\hat B_{\tS})} \!\otimes\! \hat A_{\tS}\hat B_{\tS}.
\end{align}
Using the identity
\begin{align}
  (-)^{\deg(\hat A_{\tS}\hat B_{\tS})} = (-)^{\deg\hat A_{\tS} + \deg\hat B_{\tS}},
\end{align}
and discussing case by case according to the parity of $\hat A_{\tS}$ and $\hat G_{\B}$, one is easy to show the equality.

\section{General Wick's theorem for fermionic operators}

This section presents the general Wick's theorem for fermionic operators, which will be readily used in the next Appendix. The Wick's theorem generally discusses the relation between two orderings of a set of operators.\cite{Fer21052209} We define the ordering as
\begin{align}
  \mathcal O(\hat\phi_1\cdots\hat\phi_n) = (-1)^{\#_{\mathscr p_n}}\hat\phi_{\mathscr p_1}\cdots\hat\phi_{\mathscr p_n}.
\end{align}
Here, $\#_{\mathscr p_n}$ is the number of permutations to arrange the sequence $\{1,2,\cdots,n\}$ into the sequence $\{\mathscr p_1,\mathscr p_2,\cdots,\mathscr p_n\}$. We denote the ordered label as $\mathscr p_1 \succ \mathscr p_2 \succ\cdots\succ \mathscr p_n$. Denote the linear transformation 
\begin{align}
  \hat\phi_{\alpha} = g_{\alpha k}\hat\varphi_k.
\end{align}
And the other ordering $\mathcal O'$ is defined on $\{\hat\varphi_k\}$. The core quantity is the contraction,
\begin{align}\label{c3}
  \mathcal C_{\alpha\alpha'} \equiv (\mathcal O - \mathcal O')(\hat\phi_\alpha\hat\phi_{\alpha'}) = (\theta_{\alpha\succ\alpha'} - \theta_{k\succ k'})\{\hat\phi_{\alpha},\hat\phi_{\alpha'}\},
\end{align}
where we introduce the ordering step function, $\theta_{\alpha\succ\alpha'}$ with $\theta_{\alpha\succ\alpha'} = 1$ if $\alpha\succ\alpha'$ and $\theta_{\alpha\succ\alpha'} = 0$ otherwise. And $\theta_{k\succ k'}$ is similar. Note that the second term should be explained as 
\begin{align}
  \theta_{k\succ k'}\{\hat\phi_{\alpha},\hat\phi_{\alpha'}\} \equiv \sum_{kk'}\theta_{k\succ k'}g_{\alpha k}g_{\alpha' k'}\{\hat\varphi_k,\hat\varphi_{k'}\}.
\end{align}
The general Wick's theorem states that if all $\{\hat\phi_{\alpha},\hat\phi_{\alpha'}\}$ are c-numbers, changing the ordering from $\mathcal O$ to $\mathcal O'$ gives \cite{Fer21052209}
\begin{align}\label{c4}
  \mathcal O\prod_{i=1}^n\hat\phi_{\alpha_i} = \mathcal O'\prod_{i=1}^n\hat\phi'_{\alpha_i}
\end{align}
with 
\begin{align}\label{c5}
  \hat\phi'_\alpha \equiv \hat\phi_{\alpha} + \sum_{\alpha'}\mathcal C_{\alpha\alpha'}\partial_{\alpha'}.
\end{align}
Here, $\partial_{\alpha'}$ is the Grassmann derivative with respect to $\hat\phi_{\alpha'}$, satisfying 
\begin{align}
  \{\partial_{\alpha}, \partial_{\alpha'}\} = 0\text{ and }\{\partial_{\alpha}, \hat\phi_{\alpha'}\} = \delta_{\alpha\alpha'}.
\end{align}

We prove the theorem by induction. Firstly, the cases that $n=0$ and $n=1$ are trivial. Assume the theorem holds for $n$. We will show that 
\begin{align}
  \mathcal O\hat\phi_\alpha\prod_{i=1}^n\hat\phi_{\alpha_i} = \mathcal O'\hat\phi'_\alpha\prod_{i=1}^n\hat\phi'_{\alpha_i}.
\end{align}
Assume that $\alpha_n\succ\cdots\succ\alpha_1$, and we have 
\begin{align}
  \mathcal O\prod_{i=1}^n\hat\phi_{\alpha_i} = \hat\phi_{\alpha_n}\cdots\hat\phi_{\alpha_1}.
\end{align}
Then, 
\begin{align}
  \mathcal O\hat\phi_\alpha\prod_{i=1}^n\hat\phi_{\alpha_i} = (-)^{n-j}\hat\phi_{\alpha_n}\cdots\hat\phi_{\alpha_{j+1}}\hat\phi_\alpha\hat\phi_{\alpha_j}\cdots\hat\phi_{\alpha_1}.
\end{align}
To apply the case $n$, we move $\hat\phi_\alpha$ to the leftmost side. The first move gives
\begin{align}
  &\quad\ \hat\phi_{\alpha_n}\cdots\hat\phi_{\alpha_{j+1}}\hat\phi_\alpha\hat\phi_{\alpha_j}\cdots\hat\phi_{\alpha_1}\nl
  & = -\hat\phi_{\alpha_n}\cdots\hat\phi_\alpha\hat\phi_{\alpha_{j+1}}\hat\phi_{\alpha_j}\cdots\hat\phi_{\alpha_1} \nl
  &\quad\, + \{\hat\phi_{\alpha},\hat\phi_{\alpha_{j+1}}\}\hat\phi_{\alpha_n}\cdots\hat\phi_{\alpha_{j+2}}\hat\phi_{\alpha_j}\cdots\hat\phi_{\alpha_1}\nl
  & = -\hat\phi_{\alpha_n}\cdots\hat\phi_\alpha\hat\phi_{\alpha_{j+1}}\hat\phi_{\alpha_j}\cdots\hat\phi_{\alpha_1} \nl
  &\quad\, - (-)^{n-j}\{\hat\phi_{\alpha}, \hat\phi_{\alpha_{j+1}}\} \partial_{\alpha_{j+1}}\hat O\prod_{i=1}^n\hat\phi_{\alpha_i}.
\end{align}
Iteratively, 
\begin{align}
  \mathcal O\hat\phi_\alpha\prod_{i=1}^n\hat\phi_{\alpha_i} = \hat\phi_{\alpha}\mathcal O\prod_{i=1}^n\hat\phi_{\alpha_i} - \sum_{\alpha'\succ\alpha}\{\hat\phi_{\alpha},\hat\phi_{\alpha'}\}\partial_{\alpha'}\mathcal O\prod_{i=1}^n\hat\phi_{\alpha_i}.
\end{align}
Using \Eq{c5} and the induction assumption, we obtain 
\begin{align}\label{c12}
  \mathcal O\hat\phi_\alpha\prod_{i=1}^n\hat\phi_{\alpha_i} &= \hat\phi_{\alpha}'\mathcal O'\prod_{i=1}^n\hat\phi'_{\alpha_i} - \sum_{\alpha'}\mathcal C_{\alpha\alpha'}\partial_{\alpha'}\mathcal O'\prod_{i=1}^n\hat\phi'_{\alpha_i} \nl
  &\quad\, -\sum_{\alpha'\succ\alpha}\{\hat\phi_{\alpha},\hat\phi_{\alpha'}\}\partial_{\alpha'}\mathcal O'\prod_{i=1}^n\hat\phi'_{\alpha_i}.
\end{align}
Define 
\begin{align}
  \hat\phi_\alpha' \equiv g_{\alpha k}\hat\varphi_k' 
\end{align}
with 
\begin{align}
  \hat\varphi_k' \equiv \hat\varphi_k + \sum_{k'}\mathcal C_{kk'}\tilde\partial_{k'}.
\end{align}
Here, the derivative is $\tilde\partial_{k} \equiv {\partial}/{\partial\hat\varphi_k}$ and the contraction is defined as $\mathcal C_{kk'}\equiv \sum_{\alpha\alpha'}g_{\alpha k}g_{\alpha k'}\mathcal C_{\alpha\alpha'}$. It is easy to verify 
\begin{align}
  \{\hat\varphi_{k}', \hat\varphi_{k'}'\} = \{\hat\varphi_k, \hat\varphi_{k'}\}.
\end{align}
Then we have 
\begin{align}
  \mathcal O'\hat\varphi'_k\prod_{i=1}^n\hat\varphi'_{k_i} = \hat\varphi'_{k}\mathcal O'\prod_{i=1}^n\hat\varphi'_{k_i}\! -\! \sum_{k'\succ k}\{\hat\varphi_{k}',\hat\varphi_{k'}'\}\tilde\partial_{k'}\mathcal O'\prod_{i=1}^n\hat\varphi'_{k_i}.
\end{align}
Consequently, 
\begin{align}\label{c17}
  \hat\phi'_{\alpha}\mathcal O'\prod_{i=1}^n\hat\phi'_{\alpha_i} &= \mathcal O'\hat\phi'_{\alpha}\prod_{i=1}^n\hat\phi'_{\alpha_i} \nl
  &\quad\, + \sum_{k'\succ k}g_{\alpha k}\{\hat\varphi_k,\hat\varphi_{k'}\}\tilde\partial_{k'}\mathcal O'\prod_{i=1}^n\hat\phi'_{\alpha_i}.
\end{align}
Substituting \Eq{c17} into \Eq{c12} and using \Eq{c3}, we arrive at the desired result.

\section{Wick's theorem for fermionic superoperators}

\subsection{Wick's theorem for any ordering}

In this Appendix, we simply review Wick's theorem for fermionic superoperators. The Wick's theorem is applied to reduce the multi-point average over the thermal state of fermionic operators to a two-point one. Here, we discuss the general case with arbitrary ordering of the involved operators.

Generally, we are interested in evaluating 
\begin{align}\label{Sn}
  {\rm Tr}\big( \mathscr{c}_{k_1}^{\sigma_1\lambda_1}\cdots\mathscr{c}_{k_n}^{\sigma_n\lambda_n}\rho_{\rm eq} \big).
\end{align}
Here, the average is over the equilibrium state $\rho_{\rm eq} = e^{-\beta H} / Z$ with $Z = {\rm Tr}\big( e^{-\beta H} \big)$ and 
\begin{align}
  H = \sum_k \epsilon_k \hat c_k^\dagger \hat c_k.
\end{align}
And the involved superoperators are defined as 
\begin{align}
  \begin{split}
    \mathscr c_k^{\sigma >} (\cdot)\equiv (\hat c_k^{\sigma})^> (\cdot) &= \hat c_k^\sigma(\cdot), \\
    \mathscr c_k^{\sigma <} (\cdot)\equiv (\hat c_k^{\sigma})^< (\cdot) &= (\cdot)\hat c_k^{\sigma},
  \end{split}
\end{align}
with $\sigma = \pm$ representing the creation/annihilation operators. In \Eq{Sn}, we use the index $\lambda$ to label the left and right action. Using the fermionic commutators, we have ($\bar\sigma \equiv -\sigma$)
\begin{align}
  \{\mathscr c_k^{\sigma>},\mathscr c_{k'}^{\sigma'>}\} = \{\mathscr c_k^{\sigma<},\mathscr c_{k'}^{\sigma'<}\} = \delta_{kk'}\delta^{\sigma\bar\sigma'}
\end{align}
and 
\begin{align}
  [\mathscr c_k^{\sigma >}, \mathscr c_{k'}^{\sigma' <}] = 0.
\end{align}
To proceed, we denote \cite{Sap12235432}
\begin{align}
  \mathscr j_{k}^{\sigma >} \equiv \frac{\mathscr c_k^{\sigma>} - \mathscr P_{}\mathscr c_k^{\sigma <}}{\sqrt 2},\ \mathscr j_{k}^{\sigma <} \equiv \frac{\mathscr c_k^{\sigma>} + \mathscr P_{}\mathscr c_k^{\sigma <}}{\sqrt 2},
\end{align}
with $\mathscr P(\cdot) \equiv \hat P(\cdot)\hat P$ and $\hat P \equiv \prod_k\exp(i\pi\hat c_k^\dagger\hat c_k)$. The corresponding anti-commutation relations read 
\begin{align}
  \{\mathscr j_k^{\sigma>}, \mathscr j_{k'}^{\sigma'<} \} = \{ \mathscr j_k^{\sigma<}, \mathscr j_{k'}^{\sigma'>} \} = \delta_{kk'}\delta^{\sigma\bar\sigma'}
\end{align}
and 
\begin{align}
  \{\mathscr j_k^{\sigma>}, \mathscr j_{k'}^{\sigma'>} \} = \{ \mathscr j_k^{\sigma<}, \mathscr j_{k'}^{\sigma'<} \} = 0,
\end{align}
which resemble the conventional fermionic algebra. For simplicity, we denote $\lambda=1$ for $>$ and $\lambda=-1$ for $<$, and $\bar \lambda \equiv -\lambda$. Then we have 
\begin{align}\label{jj}
  \{\mathscr j_k^{\sigma \lambda}, \mathscr j_{k'}^{\sigma'\lambda'}\} = \delta_{kk'}\delta^{\sigma\bar\sigma'}\delta^{\lambda\bar \lambda'}.
\end{align}

For presenting the Wick's theorem, we define a certain ordering of a set of $\{\mathscr j_{k}^{\sigma p}\}$, 
\begin{align}\label{Op}
  \mathcal O (\mathscr j_{k_1}^{\sigma_1 \lambda_1}\cdots\mathscr j_{k_n}^{\sigma_n \lambda_n}) \equiv (-1)^{\#_{\mathscr p_n}}\mathscr j_{k_{\mathscr p_1}}^{\sigma_{\mathscr p_1}\lambda_{\mathscr p_1}}\cdots \mathscr j_{k_{\mathscr p_n}}^{\sigma_{\mathscr p_n}\lambda_{\mathscr p_n}}.
\end{align}
Here, $\#_{\mathscr p_n}$ is the number of permutations to arrange the sequence $\{1,2,\cdots,n\}$ into the ordered one $\{\mathscr p_1,\mathscr p_2,\cdots,\mathscr p_n\}$ with $\mathscr p_1\succ \mathscr p_2 \succ \cdots \succ \mathscr p_n$. Thus, the Wick's theorem reads (for $n$ even) \cite{Sap12235432}
\begin{align}\label{d11}
  &\quad\ {\rm Tr}\big( \mathcal O\mathscr j_{k_1}^{\sigma_1\lambda_1}\cdots\mathscr j_{k_n}^{\sigma_n\lambda_n}\rho_{\rm eq} \big)\nl
  & = \sum_{\mathscr p \in \mathcal P_n} (-)^{\#_\mathscr p}\!\!\prod_{(i, j)\in \mathscr p}\!\!\!{\rm Tr}(\mathcal O\mathscr j_{k_i}^{\sigma_i\lambda_i}\mathscr j_{k_j}^{\sigma_j\lambda_j}\rho_{\rm eq}).
\end{align}
Here, $\mathcal P_n$ is the set of all possible ordered pairs $(i,j)$ with $i\succ j$ and $(-1)^{\#_\mathscr p}$ is the sign for permuting the original order into the order $\mathscr p$. For $n$ odd, the average is zero.

\subsection{Proof}

We focus on the quantity 
\begin{align}
  S_n \equiv {\rm Tr}\big( \mathcal O\mathscr j_{k_1}^{\sigma_1\lambda_1}\cdots\mathscr j_{k_n}^{\sigma_n\lambda_n}\rho_{\rm eq} \big).
\end{align}
Note that for an operator $\hat O$,
\begin{align}\label{>O}
  {\rm Tr}(\mathscr j_{k}^{\sigma >}\hat O) = \frac{1}{\sqrt 2}{\rm Tr}\big( \hat c_k^\sigma\hat O - \hat P\hat O\hat c_k^\sigma\hat P \big) = 0.
\end{align}
Thus we denote the normal ordering $\mathcal N$ for superoperators by putting all $\mathscr j_k^{\sigma >}$ to the right side of $\mathscr j_k^{\sigma <}$. And for $\sigma$ and $k$, we order them by firstly arranging $\sigma = +$ to the left side of $\sigma = -$, and then arranging $k$ in an ascending order. For example, we have
\begin{align}
  \mathcal N(\mathscr j_{k_1}^{+>}\mathscr j_{k_2}^{-<}\mathscr j_{k_3}^{+<}\mathscr j_{k_4}^{->}) = - \mathscr j_{k_1}^{+>}\mathscr j_{k_4}^{->}\mathscr j_{k_3}^{+<}\mathscr j_{k_2}^{-<},
\end{align}
where the minus sign is from odd number of permutations. From \Eq{>O}, we know that 
\begin{align}\label{d15}
  {\rm Tr}\big(\mathcal N\mathscr j_{k_1}^{\sigma_1\lambda_1}\cdots\mathscr j_{k_n}^{\sigma_n\lambda_n}\rho_{\rm eq} \big) = 0.
\end{align}
Since the superoperators $\{\mathscr j_k^{\sigma\lambda}\}$ satisfy the fermionic anti-commutation relation [\Eq{jj}], we can apply the general Wick's theorem directly. Using \Eqs{c4}, (\ref{d15}), and 
\begin{align}
  {\rm Tr}[(\mathcal O - \mathcal N)\mathscr j_{k_i}^{\sigma_i\lambda_i}\mathscr j_{k_j}^{\sigma_j\lambda_j}\rho_{\rm eq}] = {\rm Tr}(\mathcal O\mathscr j_{k_i}^{\sigma_i\lambda_i}\mathscr j_{k_j}^{\sigma_j\lambda_j}\rho_{\rm eq}),
\end{align}
\Eq{d11} is readily proved. Due the bath superoperators involved in \Eq{eq55} are linear combination of $\{\mathscr j_k^{\sigma\lambda}\}$, \Eq{d11} also holds for them. Let $\mathcal O$ be the time-ordering operator $\mathcal T_{\B}$, then we arrive at \Eq{eq56}.


%

\end{document}